\documentclass[aps,prx,twocolumn,amsmath,amssymb,superscriptaddress,floatfix]{revtex4-2}

\usepackage{mathtools}
\usepackage{multirow}
\usepackage{xcolor}
\usepackage{bm}
\usepackage{amsfonts,amssymb,amsmath}
\usepackage{graphicx,dcolumn,bm,xcolor,braket,slashed}
\usepackage{times} 
\usepackage{comment}
\usepackage{array}
\usepackage{textcomp}
\usepackage[normalem]{ulem}
\usepackage{dsfont}

\usepackage[title,toc,titletoc,page]{appendix}

\newcommand{\be}{\begin{eqnarray}}
\newcommand{\ee}{\end{eqnarray}}
\newcommand{\bbm}{\begin{bmatrix}}
\newcommand{\ebm}{\end{bmatrix}}
\newcommand{\bpm}{\begin{pmatrix}}
\newcommand{\epm}{\end{pmatrix}}

\begin{document}

\title{Third-harmonic generation in superconductors: Role of quantum geometry in the competition between Higgs mode and quasiparticles}

\author{Chang-geun Oh}
\email{cg.oh.0404@gmail.com}
\affiliation{Department of Applied Physics, The University of Tokyo, Tokyo 113-8656, Japan}
\author{Haruki Watanabe}
\affiliation{Department of Applied Physics, The University of Tokyo, Tokyo 113-8656, Japan}
\affiliation{Department of Physics, The Hong Kong University of Science and Technology, Clear Water Bay, Hong Kong, China}
\affiliation{Institute for Advanced Study, The Hong Kong University of Science and Technology, Clear Water Bay, Hong Kong, China}
\affiliation{Center for Theoretical Condensed Matter Physics, The Hong Kong University of Science and Technology, Clear Water Bay, Hong Kong, China}
\author{Naoto Tsuji}
\affiliation{Department of Physics, University of Tokyo, Bunkyo-ku, Tokyo 113-8656, Japan}
\affiliation{RIKEN Center for Emergent Matter Science (CEMS), Wako, Saitama 351-0198, Japan}
\affiliation{Trans-scale Quantum Science Institute, University of Tokyo, Bunkyo-ku, Tokyo 113-8656, Japan}


\begin{abstract}
Collective modes in superconductors, such as the Higgs mode, offer deep insights into the nature of condensates. Third-harmonic generation (THG) is a primary tool for probing the Higgs mode, but its signal competes with that of quasiparticle excitations depending on impurity scattering rates.
In particular, in the clean regime the standard BCS theory generally predicts the dominance of quasiparticle contributions.
Here, we propose and demonstrate that the quantum geometry of electronic bands can be a key mechanism governing this competition. By developing a formalism that explicitly incorporates the quantum metric, and applying it to a tunable model of a dispersive-band superconductor, we show that the quantum metric can dramatically amplify the nonlinear light-Higgs coupling by several orders of magnitude. Our results establish that a large quantum metric can cause the Higgs mode to dominate the THG response, 
resolving the puzzle of Higgs and quasiparticle competition in the clean regime and identifying band geometry as a crucial ingredient for designing and understanding the nonlinear response of superconductors.
\end{abstract}

\maketitle

\textit{Introduction.}
Superconductors are not merely perfect conductors but represent a macroscopic quantum state formed by the condensation of Cooper pairs~\cite{tinkham2004introduction,bardeen1957theory}. As a consequence of this phase transition,
superconductors host unique collective excitation modes, such as the amplitude fluctuation known as the Higgs mode and the phase fluctuation known as the Nambu-Goldstone mode~\cite{anderson1958random, Schmid1968, pekker2015amplitude, shimano2020higgs, Tsuji-Ency2024}. In charged systems, the phase mode is pushed to a high-energy scale, corresponding to the plasma frequency, by the Anderson-Higgs mechanism~\cite{anderson1963plasmons, Englert1964, Higgs1964, Guralnik1964}. In contrast, the Higgs mode remains a gapped excitation near the superconducting gap energy, $2\Delta$. Since the Higgs mode does not linearly couple to external electromagnetic fields under general circumstances, its observation has primarily relied on nonlinear response techniques, such as Raman scattering~\cite{SooryakumarKlein1980, Measson2014, Grasset2019}, third-harmonic generation (THG)~\cite{matsunaga2014light, matsunaga2017polarization, Chu2020, Kovalev2021, Wang2022, Kim2024}, and pump-probe spectroscopy~\cite{Matsunaga2013, matsunaga2014light, Katsumi2018, Vaswani2021}.

The THG signal, which appears under the resonance condition $2\Omega = 2\Delta$ (where $\Omega$ is the frequency of the incident light), has been considered a powerful tool for probing the existence of the Higgs mode~\cite{matsunaga2014light, tsuji2015theory}. However, this energy regime is also the threshold for the pair-breaking process of quasiparticles, making it a significant challenge to distinguish between the two contributions. According to the standard BCS theory for single-band superconductors in the clean limit, the THG signal is expected to be predominantly governed by quasiparticle excitations~\cite{cea2016nonlinear}.

However, experiments on materials such as NbN have indicated that the Higgs mode in fact dominates the THG resonance; this is supported by polarization-angle dependence measurements of THG~\cite{matsunaga2017polarization} and a comparison to first-principles-based calculations including impurity scattering~\cite{tsuji2020higgs}. In fact, the contribution of the Higgs mode to THG is strongly enhanced with the help of impurities~\cite{Jujo2018, murotani2019nonlinear,silaev2019nonlinear,tsuji2020higgs,seibold2021third,Derendorf2024} or phonon retardation effects~\cite{Tsuji2016}, since the paramagnetic coupling to electromagnetic fields is allowed in those situations.
While this mechanism is plausible, the interpretation of experimental results often requires a detailed comparison to material-specific theoretical calculations. Thus, a universal, material-independent principle governing the competition between the Higgs and quasiparticle channels has remained a key open question in the field~\cite{2DCS}.

Recently, the quantum geometry of electronic wavefunctions has emerged as a crucial factor in understanding various phenomena in condensed matter\,\cite{peotta2015superfluidity, torma2022superconductivity, yu2024quantum,torma2023essay}. 
The geometry of quantum states is described by the quantum geometric tensor, whose symmetric real part is the quantum metric, and whose antisymmetric imaginary part is the Berry curvature\,\cite{provost1980riemannian, ma2010abelian, berry1989quantum}. 
While the role of the Berry curvature is well-established as a cornerstone of topological physics\,\cite{nagaosa2010anomalous, Xiao_2010_RMP}, the physical consequences of the quantum metric have only recently begun to be explored in depth, with significant implications for superfluidity\,\cite{torma2022superconductivity, peotta2015superfluidity, liang2017band, verma2024geometric,HuPRB2022,torma2018quantum, iskin2024coherence, iskin2025pair,tian2023evidence}, optical responses\,\cite{oh2025universal,oh2025color,cook2017design, de2017quantized, bhalla2022resonant,ghosh2024probing, ezawa2024analytic,ahn2022riemannian, chen2025dielectric,ezawa2025quantum}, magnetic properties\,\cite{oh2025magnetic,oh2024bilayer, kitamura2024spin}, and other electronic properties \cite{rhim2020quantum,hwang2021geometric, panahiyan2020fidelity, oh2022bulk,kim2023general, oh2024thermoelectric}. The profound impact of quantum geometry on superconductivity has been particularly highlighted in the context of flat-band systems, where the quantum metric, rather than band dispersion, governs the superfluid weight \cite{torma2022superconductivity, peotta2015superfluidity, liang2017band,tian2023evidence}. 
While a recent study~\cite{xiao2025effects} showed that quantum geometry leads to Higgs-mode dominance in THG response in flat-band systems, a universal mechanism that governs the competition between Higgs and quasiparticle channels in general dispersive-band superconductors—where standard BCS theory traditionally predicts quasiparticle dominance—has yet to be fully elucidated.


In this paper, we show that the quantum geometry of the electronic wavefunction plays a key role in determining the relative contributions of the Higgs mode and quasiparticles in the THG response. We  demonstrate that the light-matter interaction mediated by this quantum metric acts asymmetrically on the Higgs mode and quasiparticle channels, leading to a dramatic amplification of the nonlinear optical response of the Higgs mode.

\textit{Formalism.}
We begin with a Hamiltonian for a multiband system (setting $\hbar=e=c=1$),
\begin{align}
\mathcal H &= \mathcal H_0 + \mathcal H_{\rm int},\\
\mathcal H_0 &= \sum_{\bm k\sigma\alpha\beta} c^\dagger_{\bm k\alpha\sigma}\,\mathcal H_{\alpha\beta}^\sigma(\bm k)\, c_{\bm k\beta\sigma},\\
\mathcal H_{\rm int} &=-\sum_{\bm k\bm k'\alpha\beta} U_{\alpha\beta}\,
c^\dagger_{\bm k\alpha\uparrow}c^\dagger_{-\bm k\beta\downarrow}
c_{-\bm k'\beta\downarrow}c_{\bm k'\alpha\uparrow}.
\end{align}
Introducing the path integral and decoupling the interaction in the pairing channel by a Hubbard--Stratonovich field $\Delta_{\alpha\beta}(\tau)$ yields the action $S[c^\dagger,c,\Delta^*,\Delta]$; we defer standard steps to the Supplemental Material~\cite{SM}.

Minimal coupling to an external spatially uniform vector potential $\bm A(\tau)$ is implemented by $\bm k\!\to\!\bm k-\bm A(\tau)$ in $\mathcal H_0$. Note that for nonlocal electron interactions, the pairing function itself can couple to the gauge field~\cite{oh2024revisitingSC}. 
Assuming a local pairing interaction, only the kinetic term is modified by the vector potential.
Diagonalizing $\mathcal H_0$ at each $\bm k$ provides band eigenstates $\ket{u_{l\sigma}(\bm k)}$ and dispersions $\epsilon_{l\sigma}(\bm k)$. We assume (i) time-reversal and inversion symmetry for the normal-state band, (ii) a single band (labeled $m$, which we drop hereafter) dominates at the Fermi level, and (iii) uniform intra-orbital pairing $\Delta_{\alpha\beta}=\Delta\delta_{\alpha\beta}$, a key condition that renders the quantum geometric contribution manifest~\cite{torma2018quantum,torma2022superconductivity, peotta2015superfluidity}. Under these assumptions we project onto the single active band and obtain the effective low-energy action
\begin{align}
S &= \int_0^\beta d\tau\Big\{ N_b\frac{|\Delta|^2}{U}
+ \sum_{\bm k,\sigma} c^\dagger_{\sigma}(\bm k)[\partial_\tau+\epsilon_\sigma(\bm k-\bm A)-\mu]c_{\sigma}(\bm k)\notag\\
&\qquad\qquad -\sum_{\bm k}\big[\Delta\, f(\bm k,\bm A)\, c^\dagger_\uparrow(\bm k)c^\dagger_\downarrow(-\bm k)+\text{h.c.}\big]\Big\},
\label{eq: action}
\end{align}
where $N_b$ is the number of orbitals and 
$f(\bm k, \bm A)$ is an overlap between the eigenstates defined by
\begin{align}
f(\bm k,\bm A)=\sum_\alpha u_\alpha^*(\bm k+\bm A)u_\alpha(\bm k-\bm A)
=\braket{u_{\bm k+\bm A}|u_{\bm k-\bm A}}.
\end{align}
{Note that the vector potential appears not only in the kinetic term but also in the pairing term in Eq.~(\ref{eq: action}).
This type of coupling to gauge fields has been discussed in Refs.~\cite{Takasan2025, xiao2025effects,chen2024ginzburg}.
In conventional superconductors, the vector potential typically couples to the order parameter fluctuations indirectly through the kinetic term $\epsilon(\bm{k} -\bm{A})$. In our framework, however, the geometric form factor in Eq. (5) enables a direct coupling between the vector potential and the gap fluctuations.}
We show in the Supplemental Material~\cite{SM} that this coupling preserves gauge invariance: the gauge field and the phase mode enter only through the gauge-invariant combination $\tilde{A}_\mu=A_\mu-\partial_\mu\theta/2$, and the resulting electromagnetic kernel obeys the transverse Ward identity $q_\mu K_{\mu\nu}(\bm{q})=0$.
For a long-wavelength vector potential, we Taylor-expand $f(\bm{k},\bm{A})$. With time-reversal and inversion symmetry, the expansion reveals the quantum metric:
\begin{equation}
|f(\bm k,\bm A)| \approx 1 - 2\,g_{ij}(\bm k)\,A_iA_j + \mathcal O(A^4),
\end{equation}
with the quantum metric defined as $g_{ij}(\bm{k}) = \text{Re} \braket{\partial_i u_{\bm{k}} | (1-P_{\bm{k}}) | \partial_j u_{\bm{k}}}$, where $P_{\bm{k}}=\ket{u_{\bm{k}}}\bra{u_{\bm{k}}}$ is the projection operator onto the Bloch state.

Introducing the Nambu spinor $\psi_{\bm k}(\tau)=(c_\uparrow(\bm k),c_\downarrow^\dagger(-\bm k))^T$, the inverse Nambu Green's function reads
\begin{align}
\mathcal G^{-1}(\bm k,\tau)
&= -\partial_\tau\tau_0 -\xi_{\bm k}\tau_3 + \Delta_0\tau_1 -\Sigma(\bm k,\tau),
\end{align}
where $\xi_{\bm k}=\epsilon_{\bm k}-\mu$ and the self-energy due to amplitude fluctuation $\rho(\tau)$ and the electromagnetic field (up to $O(A^2)$) is
\begin{align}
\Sigma(\bm k,\tau)
&= \rho(\tau)\tau_1 + \frac12\partial_{ij}^2\xi_{\bm k}\,A_i(\tau)A_j(\tau)\,\tau_3 \notag\\
&\quad +2(\Delta_0+\rho(\tau))\,g_{ij}(\bm k)\,A_i(\tau)A_j(\tau)\,\tau_1.
\end{align}

Expanding the fermion determinant to quartic order in fields and integrating out the fermions and the Higgs field $\rho$ gives the effective electromagnetic action
\begin{equation}
S[A]=\sum_{ijkl}\int\!d\omega\; A_{ij}^2(-\omega)\,K_{ijkl}(\omega)\,A_{kl}^2(\omega).
\end{equation}
The optical kernel $K_{ijkl}(\omega)$ is the central result of our formalism; it decomposes into contributions from quasiparticles (qp) and the Higgs mode, $K_{ijkl} = K_{ijkl}^{\mathrm{qp}} + K_{ijkl}^{\mathrm{Higgs}}$. Crucially, each of these contributions can be further separated into a conventional term from the band dispersion (band) and a novel term from the quantum geometry (geom).

The quasiparticle contribution is explicitly decomposed as
\begin{align}
&K_{ijkl}^{\rm qp}(\omega)
= \underbrace{\sum_{\bm k}\tfrac14\partial_{ij}^2\xi_{\bm k}\partial_{kl}^2\xi_{\bm k}\chi_{33}(\bm k,\omega)}_{K^{\mathrm{qp}}_{\mathrm{band}}} \notag\\
& + \underbrace{\sum_{\bm k} 2\Delta_0\partial_{ij}^2\xi_{\bm k}g_{kl}(\bm k)\chi_{13}(\bm k,\omega)+ 4\Delta_0^2 g_{ij}(\bm k)g_{kl}(\bm k)\chi_{11}(\bm k,\omega)}_{K^{\mathrm{qp}}_{\mathrm{geom}}},
\end{align}
and the Higgs-mediated contribution reads
\begin{align}
K_{ijkl}^{\rm Higgs}(\omega)
&=-V_{ij}(\omega)\,\mathcal G_H^0(\omega)\,V_{kl}(\omega),
\end{align}
where $\mathcal{G}_H^0(\omega)$ is the bare propagator for the Higgs amplitude mode. The Higgs-light coupling vertex $V_{ij}(\omega)$ itself contains both band and geometric parts:
\begin{align}
V_{ij}(\omega) = \underbrace{\tfrac12\sum_{\bm k}\partial_{ij}^2\xi_{\bm k}\chi_{13}(\bm k,\omega)}_{V^{\text{band}}_{ij}} + \underbrace{2\Delta_0\sum_{\bm k}g_{ij}(\bm k)\chi_{11}(\bm k,\omega)}_{V^{\text{geom}}_{ij}}.
\end{align}
The bubble functions $\chi_{\alpha\beta}$ are defined in Matsubara frequency as
$\chi_{\alpha\beta}(\bm k,i\omega_m)=\frac{1}{\beta}\sum_{i\omega_n}\mathrm{Tr}\big[\mathcal G_0(\bm k,i\omega_n)\tau_\alpha\mathcal G_0(\bm k,i\omega_n+i\omega_m)\tau_\beta\big]$, with $\mathcal{G}_0$ being the mean-field Nambu Green's function. 

{Within the framework of standard BCS theory, the light-Higgs coupling is exclusively governed by the band contribution $V_{ij}^\text{band}$, which is largely suppressed in the clean limit~\cite{tsuji2020higgs, silaev2019nonlinear}.
In contrast, our geometric framework incorporates an additional intrinsic term $V_{ij}^\text{geom}$ rooted in the quantum metric.  Since the quantum metric is an intrinsic property of the electronic Bloch wavefunctions, the enhancement is expected to be robust against impurities as long as the underlying band structure remains well-defined. 
Thus, the introduction of impurities further activates paramagnetic channels~\cite{tsuji2020higgs}—effectively providing an extrinsic enhancement $V_{ij}^\text{imp}$—which then acts in synergy with the intrinsic geometric contribution to amplify the total Higgs response.}

Finally, the third-harmonic current for a monochromatic drive $A_i(t)=a_i e^{-i\Omega t}+a_i^*e^{i\Omega t}$ is obtained by functional differentiation $j_m(t) = -{\delta S}/{\delta A_m(t)}$:
\begin{align}
j_m(3\Omega)=-4\sum_{jkl}\big[K_{mjkl}^{\rm qp}(2\Omega)+K_{mjkl}^{\rm Higgs}(2\Omega)\big]\,a_ja_ka_l.
\end{align}

\textit{Application to a Quadratic Band Touching Model.}

\begin{figure}[t]
    \centering
    \includegraphics[width=\linewidth]{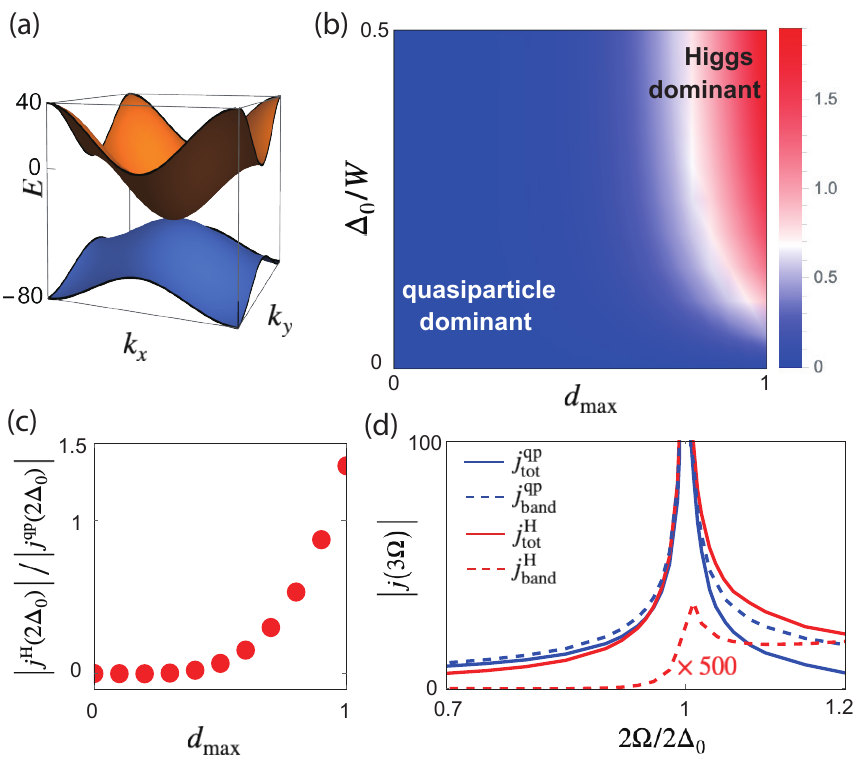} 
    \caption{
    Quantum geometric enhancement of the Higgs mode in the THG response.
    (a) Energy band dispersions of the quadratic band touching model with $t=10$ and $t_b=-5$.
        (b) 
        Crossover diagram of the THG intensity ratio $|J^{\text{H}}_\mathrm{tot}| / |J^{\text{qp}}_\mathrm{tot}|$ versus the quantum geometry parameter $d_{\mathrm{max}}$ and the gap to band width $\Delta_0/W$, where $W=8t$ is the band width of the upper band.
        (c) The ratio of the THG peak heights at resonance as a function of $d_{\mathrm{max}}$ for a fixed dispersion, quantifying the crossover from quasiparticle to Higgs dominance.
        (d)
        THG spectrum in the Higgs-dominant regime at $d_{\mathrm{max}}=1$. Solid (dashed) lines represent the total (band-only) contributions for the Higgs (red) and quasiparticle (blue) channels. The band-only Higgs contribution is magnified by $\times 500$ for visibility, highlighting that the geometric terms are responsible for the enhancement.
        Here, the results in (b-d) were calculated with $\mu=-3t$ and in (c,d) we set $\Delta_0/W=0.1$.
    }
    \label{fig1}
\end{figure}

To isolate the role of quantum geometry, we employ a specially designed two-band model that hosts a quadratic band touching at the $\Gamma$ point ($\bm{k}=0$). The explicit form and details are described in the SI.
The model is constructed to produce a familiar energy band dispersions, given by $E_+(\bm k) = -2t(\cos k_x + \cos k_y)$ and $E_-(\bm{k})=-2t_b(\cos k_x+\cos k_y)+4(t_b-t)$, as shown in Fig.~\ref{fig1}(a). A key feature of this model is that these energy dispersions remain fixed while the quantum geometry of the Bloch states is continuously tuned by a parameter $d_\mathrm{max} \in [0,1]$, which physically corresponds to the maximum quantum distance between Bloch states within the same band over the Brillouin zone and directly controls the quantum metric~\cite{rhim2020quantum,oh2022bulk,oh2025universal}.
Since all components of the quantum metric are proportional to its square ($g_{ij}\propto d_\mathrm{max}^2$), $d_\mathrm{max}=0$ defines a geometrically trivial model, while $d_\mathrm{max}=1$ represents a maximally nontrivial case.
This construction allows us to systematically isolate geometric effects on the THG response.

In our analysis, we focus exclusively on the upper electronic band, $E_+(\bm{k})$, (assuming $t_b < 0$) and consider the Fermi level sufficiently far from the band touching point at $\bm{k}=\bm{0}$. 
Provided that the superconducting gap and relevant excitation energies satisfy $\Delta_0, \hbar\Omega, k_BT \ll \text{min}_{k\in \text{FS}}|E_+(k)-E_-(k)|$, interband pairing and virtual interband transitions are negligible, and the single-band effective theory introduced in the previous section is justified.

Our central finding using this model is that the quantum geometry, tuned by the parameter $d_{\mathrm{max}}$, decisively controls the relative strength of the Higgs mode and quasiparticle contributions to the nonlinear response. This geometry-driven effect is then significantly amplified by the ratio of the superconducting gap to the bandwidth $\Delta_0/W$, where $W=8 t$ is the band width.
The key results are summarized in Fig.~\ref{fig1}.
Figure~\ref{fig1}(b) presents a crossover diagram for the ratio of the THG intensity from the Higgs mode to that from quasiparticles, $|J^{\text{H}}_\mathrm{tot}| / |J^{\text{qp}}_\mathrm{tot}|$, as a function of the quantum geometry parameter $d_{\mathrm{max}}$ and the gap $\Delta_0/W$. For this calculation, we set the chemical potential to $\mu=-3t$. 
The primary trend, observed by moving horizontally, is a transition from a quasiparticle-dominant regime (blue) at small $d_\mathrm{max}$ to a Higgs-dominant regime (red) as the geometric contribution increases. 
Crucially, the influence of quantum geometry is enhanced in systems with flatter bands.
As $\Delta_0/W$ increases (moving vertically), the crossover to Higgs dominance occurs at a smaller critical value of $d_\mathrm{max}$.
This demonstrates that a flatter band structure makes the nonlinear response more sensitive to quantum geometry, providing a clear connection to the flat-band limit ($\Delta_0/W\gg1$).
Indeed, it has been recently shown for ideal flat-band systems that the THG response is almost entirely attributed to the Higgs mode, with the quasiparticle contribution being negligible due to quantum geometric effects~\cite{xiao2025effects}.

The dramatic enhancement by geometry is quantified in Fig.~\ref{fig1}(c), which shows the ratio of the peak heights at resonance as a function of $d_{\text{max}}$ for a fixed $\Delta_0/W=0.1$. For negligible geometric effects ($d_{\mathrm{max}} \to 0$), the Higgs contribution is heavily suppressed, with the ratio being on the order of $\approx 10^{-3}$, consistent with expectations from conventional theories where quasiparticle channels dominate. However, as $d_{\mathrm{max}}$ increases, the ratio grows by several orders of magnitude, demonstrating a dramatic enhancement of the Higgs-mediated THG response driven by the quantum geometry.

The impact on the THG spectrum is shown in Fig.~\ref{fig1}(d) for $\Delta_0/W=0.1$ and $d_\mathrm{max}=1$ within the Higgs-dominant regime. The red and blue solid lines represent the total contributions from the Higgs ($j^\text{H}_{\text{tot}}$) and quasiparticle ($j^\text{qp}_{\text{tot}}$) channels, while the dashed lines show the conventional (trivial) band-only contributions ($j_{\text{band}}$). At the resonance ($2\Omega/2\Delta_0 \approx 1$), the total Higgs response is significantly larger than the total quasiparticle response. Notably, comparing the solid and dashed lines reveals that this enhancement originates almost entirely from the quantum geometric terms. The conventional band-only contribution for the Higgs mode ($j_{\text{band}}^{\text{H}}$) is extremely small and has been magnified by a factor of 500 for visibility. Therefore, our results clearly identify the quantum geometry of the Bloch states as the primary mechanism responsible for amplifying the Higgs mode's signature in the nonlinear optical response.

\textit{Correlation length.}
Beyond the electromagnetic response, the quantum geometry of the Bloch states also fundamentally determines the spatial properties of the Higgs mode, such as its correlation length $\xi_H$~\cite{xiao2025effects, iskin2024coherence,iskin2024cooper}, the characteristic spatial scale over which fluctuations in the amplitude of the superconducting order parameter remain correlated.

To investigate this, we analyze the effective action for the order parameter fluctuations in the absence of an external field ($\bm{A}=0$), focusing on the static, long-wavelength limit. As detailed in the Supplemental Material\cite{SM}, by expanding the effective action for the amplitude mode $\rho(\bm{q})$ to second order in momentum $\bm{q}$, we obtain a Ginzburg-Landau-type propagator $\braket{\rho(\bm{q})\rho(-\bm{q})} \propto (r_\rho + A_{ij}q_i q_j)^{-1}$. From this, the Higgs correlation length tensor is identified as $\xi_{H,ij}^2=A_{ij}/r_\rho$.

Similar to the optical kernel, the stiffness tensor $A_{ij}$ can be decomposed into a part from the band dispersion and a part from the quantum geometry:
\begin{align}
    \xi_{H,ij}^2 = \xi_{\text{band},ij}^2+ \xi_{\text{geom},ij}^2,
\end{align}
where the explicit form of each term is given in the Supplemental Material\cite{SM}. The geometric contribution $\xi_{\text{geom},ij}^2$ is directly related to the quantum metric $g_{ij}(\bm{k})$.

To establish the direct relationship between the total superconducting coherence length $l_{SC}$ and our derived Higgs correlation length, we evaluate the static connected correlator of the projected pair operator (see SI for detailed derivations). We find that the correlator naturally decomposes into a background (bare pair bubble) and a pole (Higgs propagation) contribution, yielding $l_{SC}^2 = l_{\text{bg}}^2 + l_{\text{pole}}^2$. Crucially, quantum geometry enhances both channels. The geometric background component, $l_{\text{bg,geom}}^2$, corresponds to the anomalous coherence length identified in \cite{hu2025anomalous}. Conversely, the pole contribution encompasses the intrinsic geometric stiffness of the Higgs mode ($\xi_{\text{geom},ij}^2$) alongside a geometric vertex correction. This decomposition demonstrates that quantum geometry amplifies the macroscopic superconducting length scale through two distinct, complementary mechanisms: the intrinsic collective amplitude fluctuations and the underlying bare condensate background.

\begin{figure}[t]
    \centering
    \includegraphics[width=\linewidth]{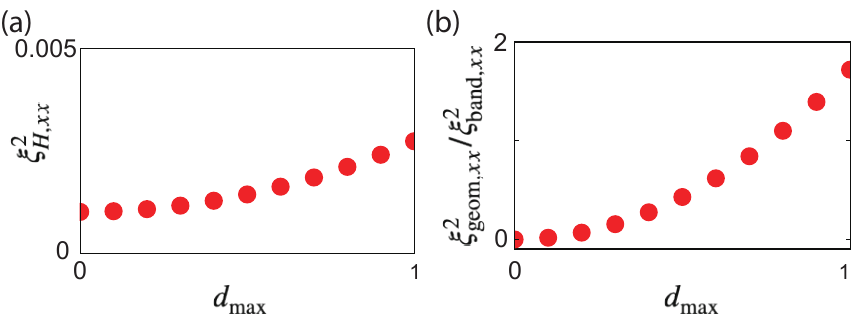}
    \caption{
        {Quantum geometric enhancement of the correlation length for the Higgs mode.} (a) The total squared correlation length of the Higgs mode, $\xi_{H,xx}^2$, as a function of the quantum geometry parameter $d_{\mathrm{max}}$ within the quadratic band touching model.
        (b) The ratio of the geometric to the band contribution to the squared correlation length, $\xi_{\text{geom},xx}^2/\xi_{\text{band},xx}^2$.
        The parameters $\mu,\Delta_0/W$ are the same as in Fig.~\ref{fig1}(b,c).}
    \label{fig2}
\end{figure}

We now quantify the geometric contribution to the correlation length using the quadratic band touching model. The calculation uses the same parameters as in the THG analysis in Fig.~\ref{fig1}(c,d). 
Since the band dispersion in our model is independent of $d_{\mathrm{max}}$, the band contribution $\xi_{\text{band},xx}^2$ remains constant. 
Therefore, any variation in the total correlation length directly measures the effect of quantum geometry. Figure~\ref{fig2}(a) shows the calculated total squared correlation length, $\xi_{H,xx}^2$, which increases monotonically as a function of $d_{\text{max}}$. To illustrate the dramatic crossover in the underlying contributions, we plot the ratio $\xi_{\text{geom},xx}^2/\xi_{\text{band},xx}^2$ in Fig.~\ref{fig2}(b). The ratio grows from zero to a value greater than one, signifying a clear transition from a band-dominated regime to one where the correlation length is overwhelmingly determined by quantum geometry.
This result unequivocally demonstrates that the quantum metric provides a significant contribution to the stiffness of the Higgs mode. While the stiffness against spatial variations of the order parameter's amplitude is primarily determined by the kinetic energy of electrons, our finding reveals that the geometric properties of the Bloch wavefunctions themselves provide an additional contribution to this stiffness. 
This enhancement of the correlation length is a direct manifestation of how the underlying geometry of the quantum states governs the macroscopic rigidity of the superconducting condensate's amplitude.

To complement the above discussion of the Higgs correlation length and the stiffness tensor $A_{ij}$, we evaluate the superfluid stiffness $D_{ij}$ for our model. 
The superfluid stiffness $D_{ij}$ decomposes into a conventional band contribution and a quantum-geometry contribution, $D_{ij} = D^{\mathrm{band}}_{ij} + D^{\mathrm{geom}}_{ij}$~\cite{liang2017band,peotta2015superfluidity}.
The geometric part increases as $d_{\max}$ increases, indicating that parameter regimes with an enhanced quantum metric not only strengthen the geometric part of $A_{ij}$ but also increase the superfluid stiffness. Detailed expressions and the $d_{\max}$-dependence are provided in SI.

\textit{Conclusion.}
In this work, we have theoretically investigated the role of quantum geometry in the nonlinear optical response of superconductors, focusing on the competition between the Higgs mode and quasiparticle contributions to THG. By developing a formalism based on a single-band projection of a multiband Hamiltonian, we explicitly derived the THG optical kernel, systematically decomposing it into band and geometric contributions.
Applying our formalism to a tunable quadratic band touching model, where the quantum geometry can be varied independently of the band structure, we have demonstrated that the quantum metric provides a significant contribution to the nonlinear light-matter coupling. Our key finding is that a prominent quantum geometry can enhance the THG response of the Higgs mode by several orders of magnitude.  Furthermore, we have shown that the Higgs correlation length is also enhanced by the quantum geometry, directly linking the spatial rigidity of the condensate's amplitude to the underlying wavefunction geometry.

Our findings provide a fundamental and general mechanism that can enhance the Higgs-mode contribution to THG.
While previous theories have invoked impurity scattering or phonon retardation effects that may depend on material details,
our work suggests that a large quantum metric can be a universal principle for achieving a strong Higgs response in dispersive-band systems in the clean regime. This shifts the focus from extrinsic factors to an intrinsic, geometric property of the electronic states themselves.

Experimentally, our work calls for a targeted search for strong Higgs responses in materials known to possess a large quantum metric, such as those with nearly flat bands. 
{Specifically, moiré superlattices, such as twisted bilayer graphene, are prime candidates as they host nearly flat bands and highly non-trivial quantum geometry~\cite{torma2022superconductivity, tian2023evidence}.
 Another important candidate is the multiband superconductor $\text{MgB}_2$, where quantum geometry plays a decisive role in the superconducting state. Notably, a recent theoretical study~\cite{yu2024non} has shown that approximately 90$~\%$ of the electron-phonon coupling constant in $\text{MgB}_2$ originates from the quantum geometric contribution of the Bloch wavefunctions.}
Probing the THG response while tuning the quantum geometry—for instance, by applying an electric field or strain—could provide a direct verification of our theory. Ultimately, our study establishes the engineering of quantum geometry as a new and promising route for controlling and amplifying the collective dynamics of superconductors.

\begin{acknowledgments}
The authors thank J. Rhim for useful discussions.
C.O. was supported by JSPS KAKENHI Grant No.~JP25KF0186.
H. W. was supported by JSPS KAKENHI grant No.~JP24K00541.
N.T. acknowledges support by JST FOREST (Grant No.~JPMJFR2131) and JSPS KAKENHI (Grant Nos.~JP24H00191, JP25H01246, JP25H01251).
\end{acknowledgments}

\bibliography{ref.bib}

\end{document}